\documentclass[
	aps,
	prl,
	showpacs,
	floatfix,
	superscriptaddress,
	onecolumn,
	preprint,
 ]{revtex4-1}
\usepackage{dcolumn,bm,textcomp}
\usepackage{mathtools}
\usepackage{amsfonts}
\usepackage{textcomp}
\usepackage{floatrow} 
\usepackage{graphicx}
\usepackage{mathptmx}
\usepackage{xspace}
\usepackage{type1cm}
\usepackage{multirow}
\usepackage{textgreek}
\usepackage{soul}[]
\usepackage{placeins}
\usepackage{setspace}
\newcommand{\LiZn}{LiZn$_2$Mo$_3$O$_8$\xspace}
\newcommand{\LiZnx}{LiZn$_{2-x}$Mo$_3$O$_8$\xspace}

\newcommand{\LiZnz}{LiZn$_{1.5}$Mo$_3$O$_8$\xspace}
\newcommand{\MoOC}{Mo$_3$O$_{13}$\xspace}
\newcommand{\MoOL}{Mo$_3$O$_{8}$\xspace}
\newcommand{\Zn}{Zn$_2$Mo$_3$O$_8$\xspace}
\newcommand{\ZnI}{ZnI$_2$\xspace}
\newcommand{\degC}{$^\textrm{o}$C\xspace}
\newcommand{\MoOH}{Mo$_3$O$_4$(OH)$_3$(H$_2$O)$_6$\xspace}
\newcommand{\Shalf}{\textit{S}~=~1/2\xspace}
\newcommand{\SOneHalf}{\textit{S}~=~1/2\xspace}

\begin{document}
\title{Electronic tunability of the frustrated triangular-lattice cluster magnet \LiZnx}
\author{J. P. Sheckelton}
 \affiliation{Department of Chemistry, The Johns Hopkins University, Baltimore, MD 21218, USA}
 \affiliation{Institute for Quantum Matter and Department of Physics and Astronomy, The Johns Hopkins University, Baltimore, MD 21218, USA}
 \affiliation{Department of Materials Science and Engineering, The Johns Hopkins University, Baltimore, Maryland 21218, USA.}
\author{J. R. Neilson}\email{Present Address: Department of Chemistry, Colorado State University, Ft. Collins, Colorado 80523, USA.}
 \affiliation{Department of Chemistry, The Johns Hopkins University, Baltimore, MD 21218, USA}
 \affiliation{Institute for Quantum Matter and Department of Physics and Astronomy, The Johns Hopkins University, Baltimore, MD 21218, USA}
\author{T. M. McQueen}\thanks{\textit{mcqueen@jhu.edu}}
 \affiliation{Department of Chemistry, The Johns Hopkins University, Baltimore, MD 21218, USA}
 \affiliation{Institute for Quantum Matter and Department of Physics and Astronomy, The Johns Hopkins University, Baltimore, MD 21218, USA}
 \affiliation{Department of Materials Science and Engineering, The Johns Hopkins University, Baltimore, Maryland 21218, USA.}

\date{October 23, 2014}

\begin{abstract}
\LiZn is an electrically insulating geometrically frustrated antiferromagnet in which inorganic \MoOC clusters each behaves as a single \Shalf unit, with the clusters arranged on a two-dimensional triangular lattice. Prior results have shown that \LiZn does not exhibit static magnetic order down to at least \textit{T}~=~0.05~K, and instead possesses a valence bond ground state. Here, we show that \LiZn can be hole doped by oxidation with I$_2$ and subsequent removal of Zn$^{2+}$ cations to access the entire range of electron count, from one to zero unpaired electrons per site on the triangular lattice. Contrary to expectations, no metallic state is induced; instead, the primary effect is to suppress the number of sites contributing to the condensed valence-bond state. Further, diffraction and pair-distribution function analysis show no evidence for local Jahn-Teller distortions or other deviations from the parent trigonal symmetry as a function of doping or temperature. Taken together, the data and density functional theory calculations indicate that removal of electrons from the magnetic layers favors Anderson localization of the resulting hole and an increase in the electrical band-gap over the formation of a metallic and superconducting state. These results put strong constraints on the chemical conditions necessary to realize metallic states from parent insulating geometrically frustrated antiferromagnets.
\end{abstract}
\maketitle

Emergent phenomena result from an ensemble of entities interacting to yield properties that appear greater than the sum of their parts. These phenomena can exist on a macroscopic scale, such as flocking birds or schooling fish, and on a microscopic scale, such as in superconducting\cite{kamihara_iron-based_2008,wu_superconductivity_1987,phelan_stacking_2013,arpino_evidence_2014}, quantum spin liquids\cite{han_fractionalized_2012,pratt_magnetic_2011}, spin ice\cite{morris_dirac_2009,castelnovo_magnetic_2008,ladak_direct_2010}, charge-density wave\cite{cottingham_dynamic_2013}, or heavy fermion\cite{steglich_superconductivity_1979,neilson_mixed-valence-driven_2012} materials. Since the discovery of high-$T_c$ superconductivity in the cuprates\cite{bednorz_possible_1986,cava_bulk_1987}, a major effort in condensed matter physics has been devoted to a rational theoretical explanation of the superconducting state. A proposed mechanism put forth by P. W. Anderson explained superconductivity in the La$_2$CuO$_4$ system\cite{anderson_resonating_1987} as arising from itinerant, resonating spin singlets, or resonating valence-bonds\cite{anderson_resonating_1973} (RVBs), as the underlying magnetic ground state to cooper pairing and superconductivity in the cuprates. A system of spins on a geometrically frustrated lattice, such as those on a two-dimensional triangular\cite{reimers_spin_1993,nakatsuji_spin_2005} or kagome\cite{han_fractionalized_2012,helton_spin_2007,hagemann_geometric_2001,williams_kmn3o2ge2o7:_2014} lattice with nearest neighbor antiferromagnetic interactions, can prevent long-range static magnetic order and harbor a RVB state. A Mott insulating RVB state that is charge doped and results in superconductivity would yield strong evidence for the RVB state as the yet unresolved theoretical explanation of high-$T_c$ superconductivity.

\begin{figure}[!htb]
\centering
  \includegraphics[width=4in]{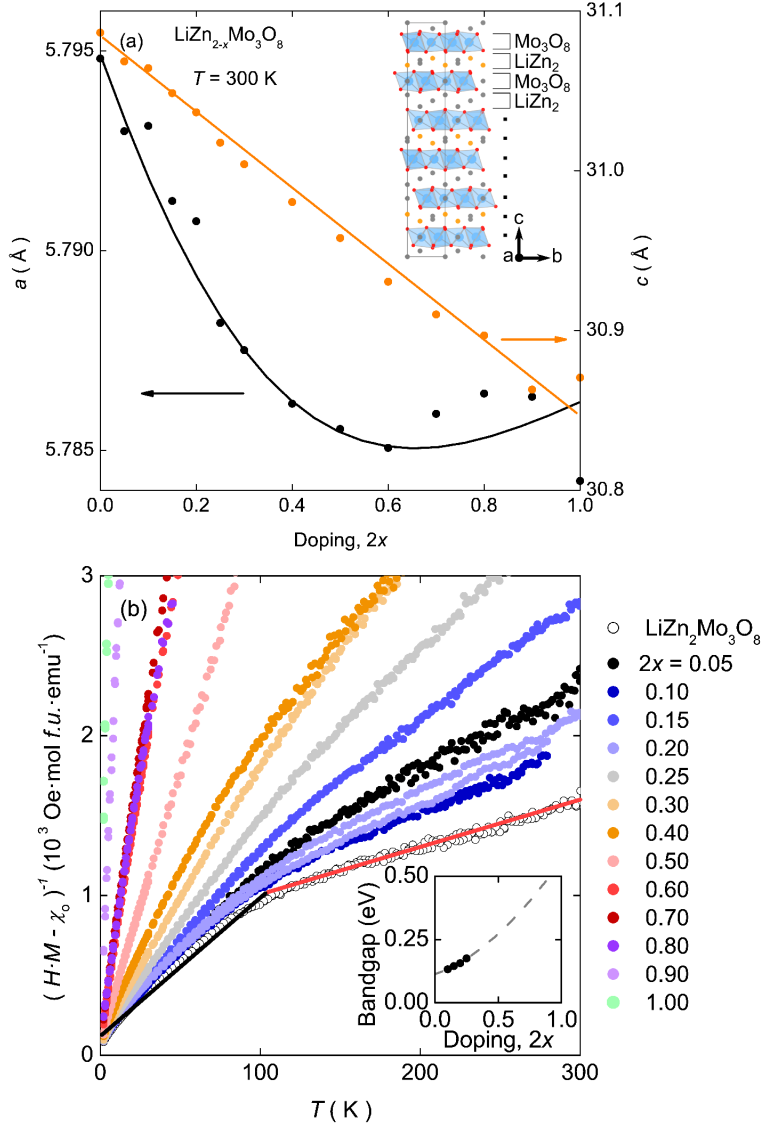}
  \caption{(a) Lattice parameters extracted from X-ray diffraction data over the entire doping range. The linear decrease in $c$ is accompanied by an flattening and possible upturn in $a$ at high doping levels. Inset: Structure of \LiZn. The Zn-site where $\sim$90\% of Zn$^{2+}$ is removed upon doping (see SI) is shown in orange. (b) Curie-Weiss analysis of the magnetic susceptibility shows that the ``kink"  at \textit{T}~=~100\,K (shown as the red and black lines on the $2x = 0$ dataset), indicative of the formation of a condensed valence-bond state, is suppressed as holes are introduced into the lattice. Inset: Electronic bandgaps extracted from the electrical resistance measurements (see SI), showing an increase upon hole doping.}
  \label{fig:MAIN_Fig_1}
\end{figure}

\LiZn is a material composed of magnetic \MoOL layers, which consist of a triangular lattice of edge-sharing \MoOC clusters, separated by alkali and alkali earth cations  (inset, Fig.~\ref{fig:MAIN_Fig_1}(a)). Each cluster has seven \textit{d}-orbital valence electrons: six forming Mo-Mo bonds and one unpaired providing an \Shalf magnetic moment \cite{sheckelton_possible_2012}. Mo-O-Mo links between \MoOC clusters provide a superexchange pathway for antiferromagnetic spin interactions, resulting in a geometrically frustrated magnetic state. Inelastic neutron scattering\cite{mourigal_molecular_2014} and a host of local magnetic measurements\cite{sheckelton_local_2014} indicate \LiZn is a spin liquid candidate, as it lacks any long range magnetic order above $T > 0.05$\,K. Bulk magnetic susceptibility measurements indicate that upon cooling \LiZn below \textit{T}~=~100\,K, two-thirds of the paramagnetic spins that previously contributed to the susceptibility effectively disappear without any experimental signature of long-range magnetic order. This is attributed to the formation of a condensed valence-bond state\cite{sheckelton_possible_2012}, in which two-thirds of the spins form valence bonds (probably resonating), while the remaining one-third spins continue to act paramagnetic down to the lowest temperatures. The origin of the stabilization of this magnetic ground state is unclear, but several theoretical proposals exist \cite{flint_emergent_2013, sheckelton_local_2014, chen_pco_2014}. 

Here we describe chemical hole doping of \LiZn over the range of one unpaired \Shalf electron per cluster to zero unpaired electrons per cluster \textit{via} oxidation of \LiZn with elemental I$_2$. These doped samples are characterized by their magnetic, transport, and structural properties \textit{via} magnetization, resistivity, X-ray powder diffraction (XRPD), neutron powder diffraction (NPD), and synchrotron X-ray powder diffraction (SXRD) experiments. Contrary to expectations, hole doping does not produce metallic behavior or superconductivity. Instead, the result is a smoothing of the transition into the condensed valence-bond state and an increase in the electrical bandgap, even though there are no average or local breaking of the trigonal symmetry. These results can be explained if the doped holes are trapped at individual sites, rather than being mobile on the lattice. Density functional theory (DFT) calculations provide a possible explanation for the lack of hole mobility in the form of a type of Anderson localization due to shifts in the individual energy levels of a \MoOC cluster when the electron count is changed from seven to six.  

Hole doping \LiZn over the range \LiZn to \LiZnz is accomplished by reacting \LiZn pellets or powder with elemental iodine\cite{rodriguez_iodine_2010} to form thermodynamically stable iodide salts with Li/Zn (see SI). Surprisingly, reaction with I$_2$ vapor results in the formation of \ZnI as opposed to the more thermodynamically stable LiI. This was confirmed by the appearance of \ZnI crystals on the cold end of the quartz reaction vessels, identified as pure \ZnI by XRPD, with no detectable LiI phase in the product \LiZnx. A possible explanation is that the higher vapor pressure of \ZnI compared to LiI allows it to be vapor transported to the cold end of the tube and irreversibly removed from the active reaction (Le Ch\^{a}telier's principle). To test this, \LiZn was reacted with I$_2$ in tetramethyene sulfone at a temperature of \textit{T}~=~150\,\degC. The reaction was incomplete, as XRPD analysis indicated the formation of two phases by splitting the $\{00l\}$ reflections. Nonetheless, a flame test revealed the presence of Li in the mothor liquor, and chemical analysis indicated that under these alternate, lower temperature, conditions both Li and Zn were being deintercalated. This result suggests that the specificity for Zn removal in the higher temperature conditions, even though Li is known to be highly mobile by NMR, is due to the irreversible removal of \ZnI by vapor transport. As it does not solely depend on relative ion mobilities in the host lattice, this method of chemical selectivity in a \textit{chimie douce} reaction, \textit{i.e.}, irreversible removal of one product from the reaction, will likely have broader application in the preparation of metastable materials.

XRPD data of all samples are well described by the trigonal $R\bar{3}m$ structure of the parent compound (see SI). The \textit{c} lattice parameter, Fig.~\ref{fig:MAIN_Fig_1}(a), follows Vegard's law and linearly decreases as a function of doping. However, the decrease is unexpected: in other systems where interlayer cations and anionic layers primarily interact \textit{via} electrostatic attraction, removal of cations causes an increase in the \textit{c} lattice parameter. The decrease indicates that in \LiZn there are more than simple electrostatic interactions between the Li/Zn layers and the \MoOL layers.  Further, even though initially decreasing, there is an unusual flattening, and possible increase, in the in-plane \textit{a} lattice parameter at high hole doping levels.

The electrical and magnetic properties also exhibit a continuous change as a function of electron count. The inverse susceptibility (see SI), Fig.~\ref{fig:MAIN_Fig_1}(b), shows a clear transition between two linear Curie-Weiss like regions in \LiZn that become less distinct as doping increases and disappear completely for $2x > 0.40$. At the same time, resistivity measurements on polycrystalline pellets, inset Fig.~\ref{fig:MAIN_Fig_1}(b), show a continuous increase in the electrical bandgap as holes are doped into \LiZnx (see SI).

\begin{figure}[!bth]
\centering
  \includegraphics[width=3in]{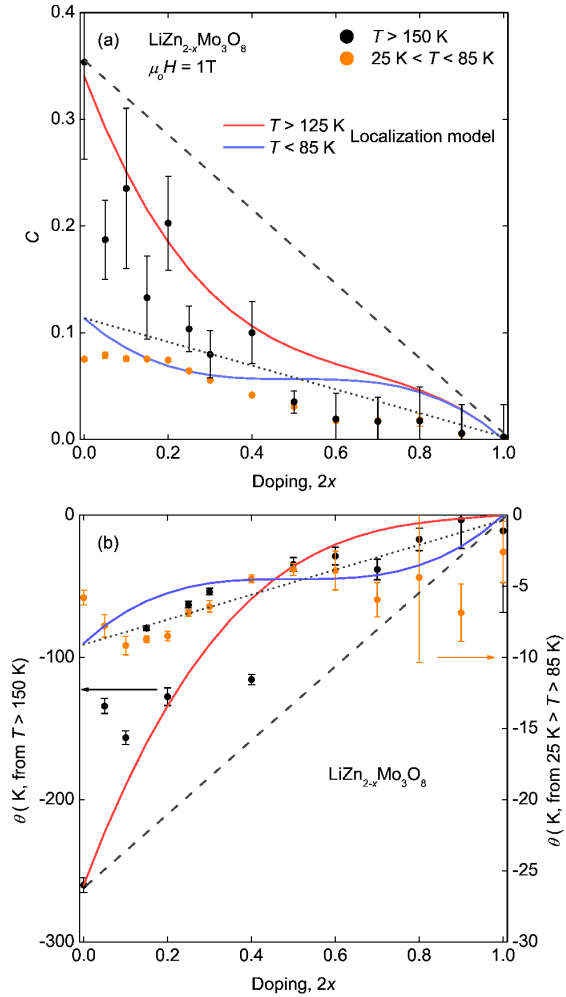}
  \caption{(a) Extracted Curie constants and (b) Weiss temperatures from two temperature ranges of the inverse susceptibility data, above \textit{T}~=~150\,K and $25\,\textrm{K}\leq T \leq 85\,\textrm{K}$. The experimental trends deviate strongly from simple linear behavior (dotted and dashed lines). Better agreement with experiment is obtained from a statistical calculation of the effect of removal of individual spins if each hole is fully localized and also localizes a neighboring valence bond (red and blue lines, see text).}
  \label{fig:MAIN_Fig_2}
\end{figure}

Extraction of the corresponding Curie constants, Fig.~\ref{fig:MAIN_Fig_2}(a), and Weiss temperatures, Fig.~\ref{fig:MAIN_Fig_2}(b), from the inverse susceptibility data reveals an interesting pattern. The Curie constant, proportional to the number of paramagnetic spins, starts close to the expected value of 0.375 for a full lattice of unpaired spins, but initially decreases three times faster than expected based on the change in electron count in the high temperature region ($T \geq 150$\,K). However, above $2x = 0.50$, further increases in hole doping changes the Curie constants at high temperature more slowly and continuously toward zero. In contrast, the Curie constants extracted from the low temperature region ($25\,\textrm{K}\leq T \leq 85\,\textrm{K}$) vary more uniformly with hole doping. Similar trends are found for the Weiss temperatures, a measure of the net magnetic interactions present, for both the high temeprature and low temperature regions.

\begin{figure}[!bth]
\centering
  \includegraphics[width=3in]{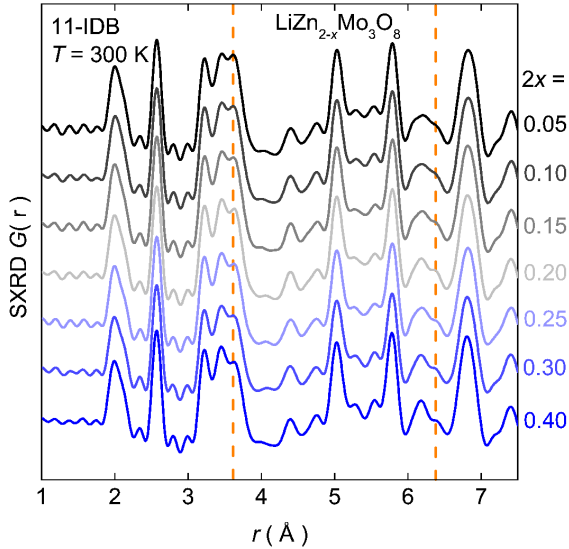}
  \caption{X-ray pair-distribution functions as a function of $r$ for doping up to $2x = 0.40$. The orange lines at $r \approx 3.6~\text{\AA}$ and $r \approx 6.3~\text{\AA}$ correspond to Zn-Mo and Zn-O distances in the average structure, which decreases as Zn is removed. There are no other significant changes in the local structure.} 
  \label{fig:MAIN_Fig_3}
\end{figure}
 
These deviations from simple expectations are not due to changes in crystallographic or local symmetry. Fig.~\ref{fig:MAIN_Fig_3} shows the pair distribution function (PDF), a  scattering factor weighted atom-atom histogram, calculated from synchrotron X-ray data at room temperature on samples up to a doping level of $2x = 0.40$. In addition to the series of nearest-neighbor metal-oxygen bonds around 2 \AA, intra- and inter-cluster Mo-Mo distances are clearly resolved at 2.6 \AA\xspace and 3.2 \AA\xspace respectively (see SI). These correlations remain sharp as holes are introduced, and rule out local structural changes of the \MoOC clusters that might explain the non-linearities in the physical properties. The only significant changes are those tied to the loss of Zn. Further, no changes are observed by neutron PDF at ambient or cryogenic conditions down to $T = 2$\,K (see SI). Rigorous analysis of synchrotron X-ray and neutron diffraction data similarly rules out large changes in the structure upon doping, beyond those expected from the change in the lattice parameters and removal of Zn (see SI).

\begin{figure}[!th]
\centering
  \includegraphics[width=3in]{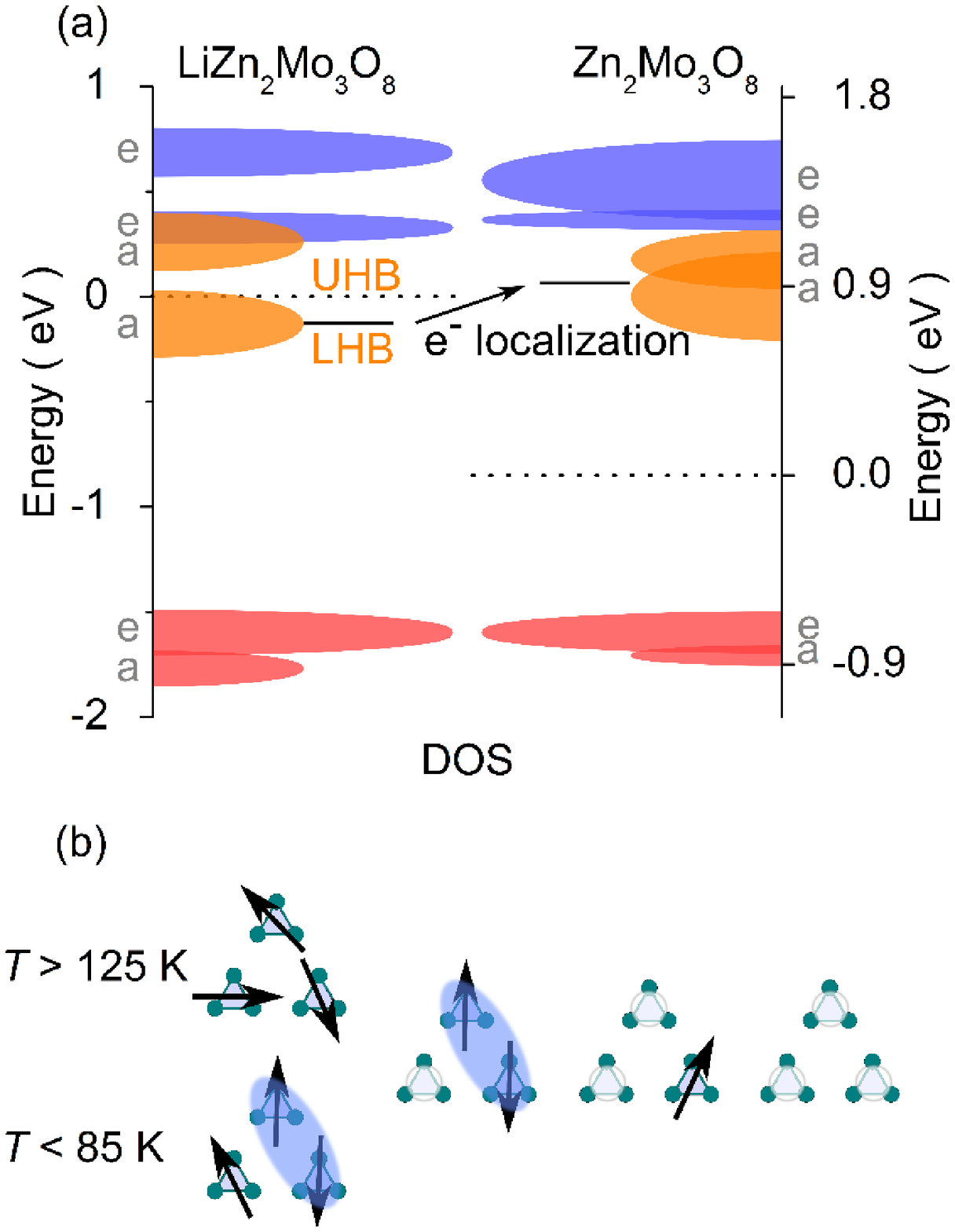}
  \caption{(a) Density of states diagram for \LiZn and \Zn. Dashed lines are respective fermi energies. The ``Zn$_2$Mo$_3$O$_8$" diagram has the same structure as \LiZn, with the Li removed. Calculations suggest the effect of hole doping this material is an increase, relative to the fermi energy and the frontier orbitals, of the band corresponding to the highest filled orbital (see SI) once an electron is removed. (b) A local model of the effect of hole doping a triangular lattice antiferromagnet: when all sites are singlely occupied (left), there is a transition from a frustrated, paramagnetic state to a condensed valence bond one where two-thirds of the electrons collapse into singlets. Upon hole doping, a distribution of local configurations is formed: one hole localizes a valence bond singlet at all temperatures, two holes liberates a free paramagnetic spin, and three holes results in a non-magnetic state.} 
  \label{fig:MAIN_Fig_4}
\end{figure}

The most likely explanation of these data is that doped holes, rather than becoming mobile in the lattice, stay strongly localized. DFT band structure calculations and the resulting density of states (DOS), Fig.~\ref{fig:MAIN_Fig_4}(a), provide insight into the origin of localization. The result of removing the magnetic valence electron from a cluster is a prominent increase in the energy of the now-empty orbital. Since the energy level is now significantly different than the filled states, it acts as a trap for the hole carrier and prevents free movement of electrons on the lattice. Put another way, the change in the electronic structure of each \MoOC cluster upon oxidation by one electron results in a form of Anderson localization\cite{anderson_absence_1958} and is confirmed by the presence of three-dimensional variable range hopping in \LiZnx resistivity measurements (see SI).

The non-linear trends in the \textit{a} lattice parameter and Curie-Weiss parameters follow from this carrier localization. Because the hole carriers are localized, the effect on the underlying magnetic lattice can be modeled by assuming that, microscopically, the frustrated lattice is built of triangles of three \MoOC clusters containing three, two, one, or zero magnetic electrons, Fig.~\ref{fig:MAIN_Fig_4}(b). The probability of occurance of each of these configurations is then set by statistics according to the fraction of holes put into the system. The easiest way to produce non-linear behavior is to assume, consistent with \LiZn being a frustrated antiferromaget, that removal of one spin from a triangle will cause the other two to ``disappear" into a singlet state. Thus, the only terms contributing to the measured susceptibility are when every site is full, and when two out of every three sites are empty. Fig.~\ref{fig:MAIN_Fig_2}(a) shows the result of calculating the high- and low-temperature Curie constants from this model (red and blue lines). The qualitative features of the data are captured: initially there is a drop of the high-temperature Curie constant three times faster than expected due to formation of singlets even at $T > 125$\,K, and then an extended plateau region where further addition of holes causes little change in the net Curie constant due to a competition between singlet localization and liberation of free spins due to two adjacent holes.  Similar logic can be used to calculate the net magnetic interaction strength and Weiss temperature, shown in Fig.~\ref{fig:MAIN_Fig_2}(b), with similar qualitative agreement with experiment that is substantially better than the na\"{\i}ve model. The agreement may be quantitative given the errors associated with the small signals being measured, but secondary effects, \textit{e.g.}, from hole (defect) clustering cannot be ruled out.

What do these results imply for the various theoretical models that have been proposed for the lack of a 120 degree magnetically ordered state in \LiZn, the expectation for a nearest-neighbor triangular lattice antiferromagnet? The diffraction and local structure measurements put strict constraints on any deviations from the trigonal symmetry, on average or locally. This would seem to rule out the previously proposed cluster rotation model \cite{flint_emergent_2013}, although small or highly dynamic (outside the energy range of the present measurements) distortions cannot be ruled out. This apparent adherence to trigonal symmetry also disfavors the explanation that the lack of magnetic order is simply due to a variation of local exchange interactions (due to Li/Zn mixing or other forms of disorder): if the frustration in \LiZn is only due to varying exchange interactions, then the magnetic ground state would be insensitive to the hole doping in \LiZnx; instead, the formation of valence bonds (singlets) appears to be strongly favored in this material near doped holes. More recently, it was proposed that \LiZn may not fall into the fully localized molecular limit and instead possess a form of partial charge order \cite{chen_pco_2014}. In this model, all of the primary effects are in the electronic channel, and any structural changes would be secondary effects resulting from the electronic changes. Qualitatively, this is in better agreement with the present results, but more thorough calculations to understand the impact of the addition of holes on the partial charge order is needed in order to make a more definitive statement.

Here we described the reactions of \LiZn with increasing amounts of I$_2$ to remove Zn$^{2+}$, thus hole doping the two dimensional triangular lattice antiferromagnet \LiZn. The effect of Zn doping is to systematically soften the transition between two linear Curie-Weiss regimes in the corrected inverse susceptibility. This doping is not accompanied by the appearance of metallic behavior or superconductivity as measured by magnetic susceptibility or resistivity. Instead, the doped holes undergo Anderson localization. This results in an increase in the electronic bandgap and the localization of neighboring singlets. Generally, these results suggest that cluster-based frustrated magnetic materials favor Anderson localization over a metal-insulator transition and/or superconductivity. It is yet to be seen if the same holds true for other (\textit{e.g.} ionic based) GFM systems. Taken together, these findings imply that realizing a metallic or superconducting state in a doped frustrated antiferromagnet may require more careful control of the local energetics than previously thought.

\section{Acknowledgments}
This research was supported by the US Department of Energy (DOE), Office of Basic Energy Sciences, Division of Materials Sciences and Engineering under Award DE-FG02-08ER46544 to The Institute for Quantum Matter at JHU. JPS would like to thank the William Hooper Grafflin Fellowship and Matt Suchomel (11-BM), Kevin Beyer (11-ID-B), and Mikhail Feygenson \mbox{(NOMAD)} for their assistance in collecting and reducing data. TMM acknowledges conversations with R. Flint, and G. Chen, and support from the the donors of the American Chemical Society Petroleum Research Fund and the David and Lucile Packard Foundation. Use of the Advanced Photon Source, an Office of Science User Facility operated for the U.S. Department of Energy (DOE) Office of Science by Argonne National Laboratory, was supported by the U.S. DOE under Contract No. DE-AC02-06CH11357. A portion of this research at ORNL's Spallation Neutron Source was sponsored by the Scientific User Facilities Division, Office of Basic Energy Sciences, U.S. Department of Energy, under contract DE-AC05-00OR22725 with UT-Battelle, LLC. 

\FloatBarrier
\newpage

\begin{center}
\large{\textbf{Supporting Information for:\\*Electronic tunability of the frustrated triangular-lattice cluster magnet \LiZnx
}}
\vspace{0.5cm}

\small{
{{J. P. Sheckelton,\textit{$^{1,2,3}$} J. R. Neilson,\textit{$^{1,2,^{\ast}}$} and T. M. McQueen\textit{$^{1,2,3,\dag}$}}}\vspace{0.5cm}

{\textit{$^{1}$~Department of Chemistry, The Johns Hopkins University, Baltimore, MD 21218, USA.}}\\
{\textit{$^{2}$~Institute for Quantum Matter, Department of Physics and Astronomy, The Johns Hopkins University, Baltimore, Maryland 21218, USA. }}\\
{\textit{$^{3}$~Department of Materials Science and Engineering, The Johns Hopkins University, Baltimore, Maryland 21218, USA. }}\\
}
\end{center}
 
\section{Experimental methods}

\LiZn was synthesized as previously reported\cite{sheckelton_possible_2012}. Polycrystalline \LiZn and stoichiometric amounts of solid I$_2$ were added to a quartz tube sealed \textit{in vacuo}, taking care to minimize loss of I$_2$ vapor from sublimation. A temperature gradient of 300\,\degC to 290\,\degC was established over the reaction vessel and held for one week, with the \LiZn/I$_2$ on the hot side. Characterization was initially performed using a Bruker D8 Focus powder X-ray diffractometer using copper K$\alpha$ ($\lambda = 1.540~\text{\AA}$) radiation and a LynxEye detector. Elemental Si was used  (\textit{a}~=~5.43102~\AA) as an internal standard for Le Bail\cite{le_bail_whole_2005} refinements. Magnetization and resistivity measurements were performed using a Quantum Design Physical Properties measurement system (PPMS). Temperature-dependent resistivity was measured on polycrystalline pellet pieces using the four-probe method over the range \textit{T}~=~300\,K to \textit{T}~=~50\,K, at which the samples became too resistive. Samples of higher doping were also too resistive to obtain meaningful numbers. Temperature-dependent magnetic susceptibility was measured on powder samples over the range \textit{T}~=~300\,K to \textit{T}~=~2\,K. Magnetization as a function of applied field was measured at \textit{T}~=~$1.8$\,K up to a field of $\mu_oH$~=~$9$\,T. Molar magnetic susceptibility, defined as $\chi_m = \partial M/\partial H$ as $H\longrightarrow0$ is approximated here as the DC magnetization over field, $M/H$. Inverse susceptibility was corrected for a temperature-independent contribution, $\chi_o$, by assuming paramagnetic behavior in the high temperature (\textit{T}~$\geq$150\,K) regime and adjusting the value of $\chi_o$ until the inverse susceptibility became linear.

High-resolution synchrotron X-ray diffraction (SXRD) experiments were carried out on powder samples of native \LiZn and $2x \leq 0.40$ doped \LiZnx samples using beamline 11-BM-B for Rietveld analysis and beamline 11-ID-B for Pair-distribution function (PDF) analysis at Argonne National Laboratory (APS)\cite{wang_dedicated_2008}. Measurements were taken on powder-coated, greased kapton capillaries at \textit{T}~=~300, 150, and 6\,K using a photon wavelength of $\lambda=0.4139(1)~\text{\AA}$ ($\approx$30~KeV) over the range $0 \leq 2\theta \leq 45$ with a 0.001$^\textrm{o}$ $2\theta$ stepsize and a measurement time of 0.1\,s per step. Diffraction scans at \textit{T}~=~6\,K were taken using an Oxford Cryosystems helium cryostat. Reitveld refinements\cite{rietveld_profile_1969} to the diffraction data were performed using the Bruker Topas professional software suite and the FullProf suite\cite{rodriguez-carvajal_recent_1993}. Powder neutron diffraction on a sample of undoped \LiZn at \textit{T}~=~300\,K  with d-spacing of $0.2760~\text{\AA} \leq d \leq 3.0906~\text{\AA}$ and $1.6557~\text{\AA} \leq d \leq 8.2415~\text{\AA}$ was performed at the POWGEN diffractometer at the Spallation Neutron Source at Oak Ridge National Laboratory (ORNL). A combined refinement to X-ray/Neutron data on the same sample of undoped \LiZn was performed to freely refine the lithium and zinc occupancies. Subsequent doped samples, with only SXRD scans, assume, based on the observation of only ZnI$_{2}$ from doping, fixed lithium occupancies  extracted from the combined X-ray/neutron refinement on \LiZn. Total scattering powder diffraction patterns for PDF analysis were collected on beamline 11-ID-B at \textit{T}~=~300\,K using a Perkin Elmer amorphous silicon detector in a flat-plate geometry and CeO$_2$ as a calibration standard. Wavelengths of $\lambda = 0.2128~\text{\AA}$ ($\approx~60~KeV$) and $\lambda = 0.1370~\text{\AA}$ ($\approx~90~KeV$) were used. Approximate sample-to-detector distances of 18 and 22 cm were used for each sample changer and precise distances were extracted from refinements to the CeO$_2$ standard. Samples were sealed in kapton tubes and diffraction patterns were collected by summing 120 scans, at a 1~second beam exposure for each scan. PDF analysis was performed on $G(r)$, extracted with a $Q_{max} = 25~\text{\AA}^{-1}$ using PDFgetX2\cite{Qiu_it_2004}. Powder neutron diffraction patterns suitable for PDF analysis were performed on undoped Li-7 isotopically enriched \LiZn at the NOMAD diffractometer at ORNL, at temperatures ranging from \textit{T}~=~2\,K to \textit{T}~=~300\,K. PDF analysis was performed on $G(r)$ extracted from data reduced using in-house scripts. Least-squares refinements to $G(r)$ were performed using PDFgui\cite{farrow_pdffit2_2007}. Refined structures were visualized with VESTA\cite{momma_<i>vesta</i>_2008}.

Molecular density functional theory (DFT) calculations were performed as previously reported\cite{sheckelton_possible_2012} using PBE0 hybrid functionals at the UHF level level of theory on the cluster Mo$_3$O$_4$(OH)$_3$(H$_2$O)$_6$, isoelectronic to a \MoOC cluster in \LiZn. The GAMESS\cite{schmidt_general_1993} software package was used for calculations and MacMolPlot\cite{bode_macmolplt:_1998} for visualization of orbitals and energies. Band structure calculations were performed on \LiZn and doped \LiZn models accounting for spin orbit coupling and a Hubbard U of 2 eV using the ELK all-electron full-potential linearized augmented-plane wave (FP-LAPW) code (Available under the GNU General Public License at elk.sourceforge.net). To simplify calculations, a \LiZn unit cell in the rhombohedral setting was used. In addition calculations were performed with fully occupied and long-range ordered Li and Zn positions in the unit cell, taking care to have the stoichiometry set to \LiZn. The doped \LiZn version is the same exact unit cell, with the Li atoms removed as to achieve a non-magnetic calculation. 

\FloatBarrier
\section{Laboratory X-ray Diffraction}

\begin{figure}[!htb]
\centering
  \includegraphics[width=4in]{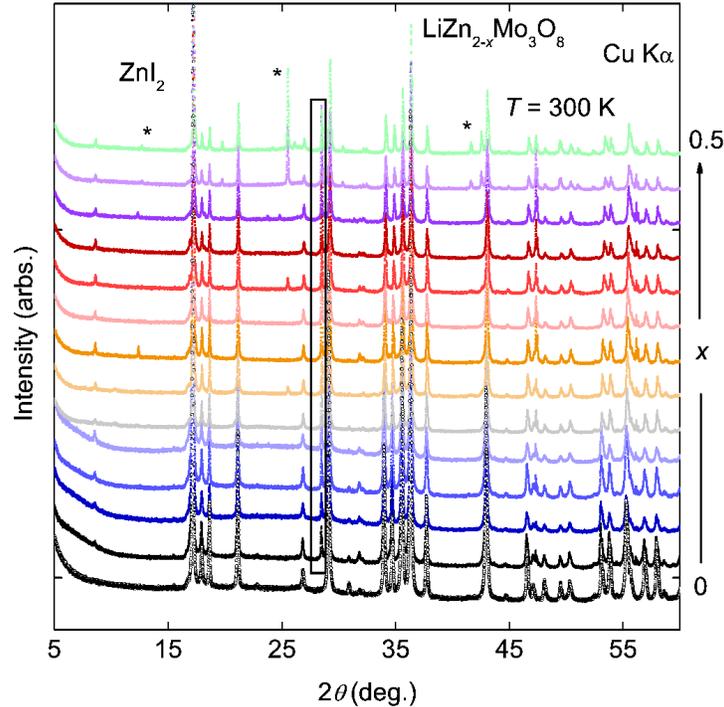}
  \caption{Laboratory powder X-ray diffraction patterns for the entire series of \LiZnx. The box highlights the strongest reflection coming from the added silicon standard. The highest doped samples show reflections identified as belonging to ZnI$_2$ (marked with *), with some samples also showing reflections identified as SiO$_2$ from breaking of the quartz ampoules.}
  \label{fig:LAB_XRD}
\end{figure}

\FloatBarrier
\newpage
\section{Magnetic Susceptibility}

\begin{figure}[!bth]
\centering
  \includegraphics[width=4.5in]{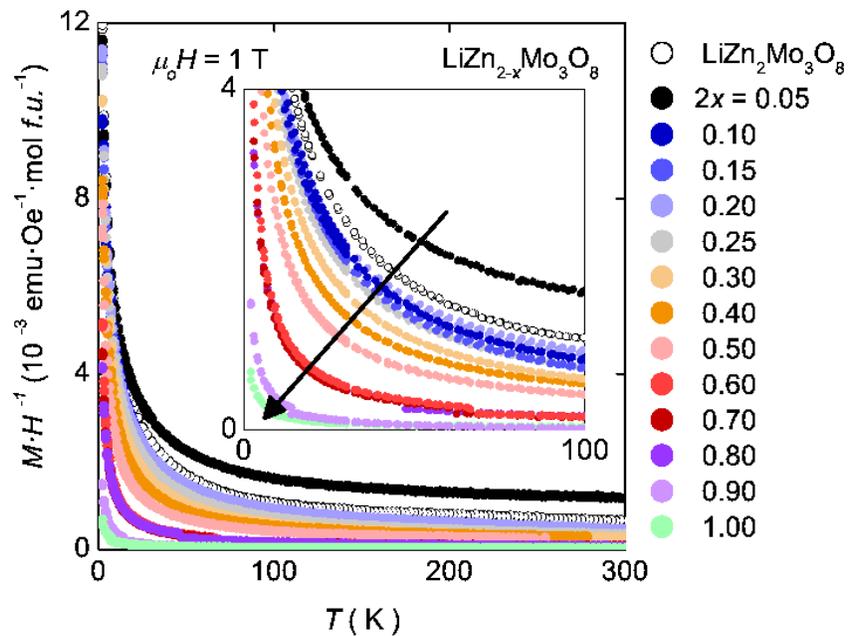}
  \caption{Magnetic susceptibility as a function of temperature for the \LiZnx series. A trend of decreasing susceptibility is observed. The small discontinuities observed in some of the datasets are experimental artifacts from the small measured signal size.}
  \label{fig:Suscept}
\end{figure}

\FloatBarrier
\newpage
\section{Resistivity}

\begin{figure}[!bht]
\centering
  \includegraphics[width=4in]{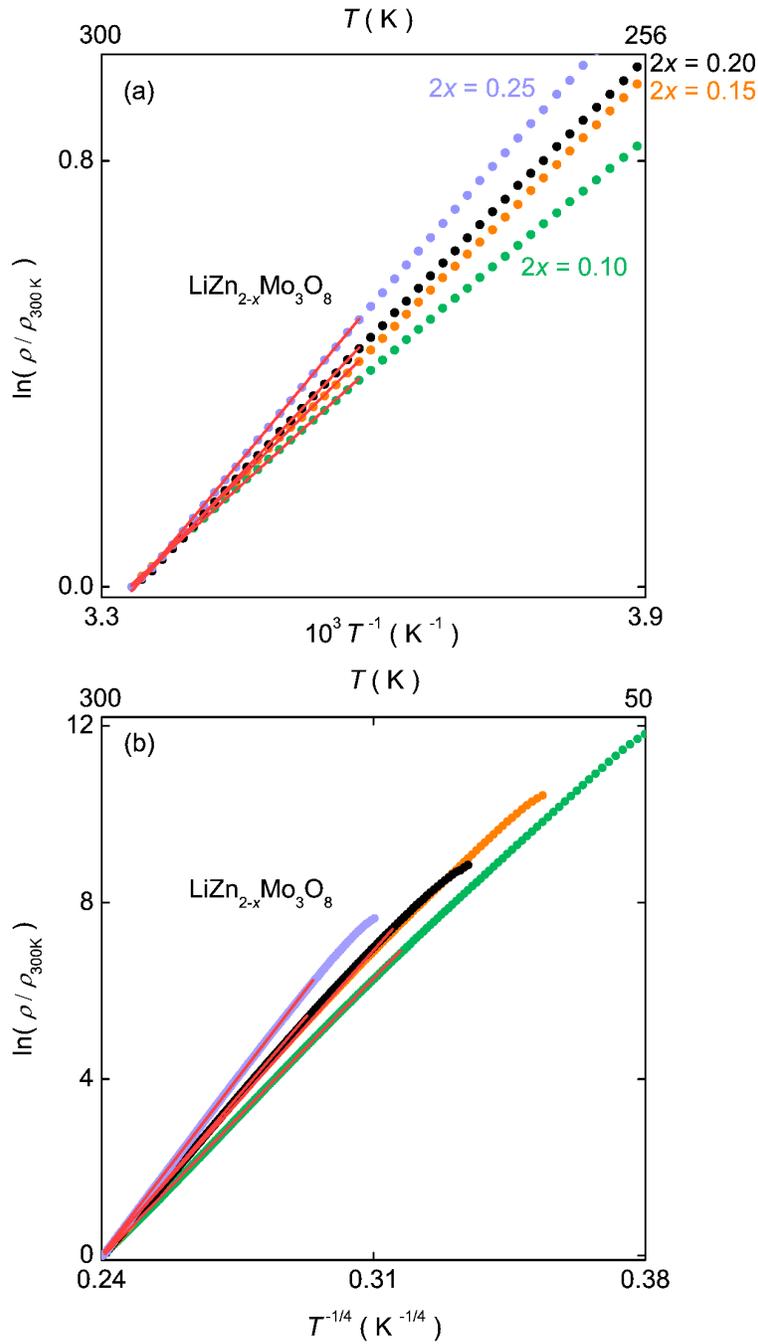}
   \caption{(a) Natural logarithm of normalized resistivity as a function of $\frac{1}{T}$ measured on pellets of \LiZnx samples (b)-(e) with fits to $ln\frac{\rho}{\rho_{300 K}}=\frac{E_g}{k_B}\frac{1}{T}$ at high temperature (red lines), used to extract the effective bandgap. (b) The same normalised resistivity in (a), but plotted vs. $\frac{1}{T^{1/4}}$. A $T^{1/4}$ scaling is suggestive of 3-dimensional variable-range-hopping (VRH) and Anderson localization. Plotting $ln\frac{\rho}{\rho_{300 K}}$ vs. $\frac{1}{T^{1/4}}$ results in a much larger range of linear behavior, indicating VRH and Anderson localization are occuring in doped samples of \LiZnx. Red lines are guides to the eye.}
  \label{fig:resist}
\end{figure}

\FloatBarrier
\section{Electronic structure}

\begin{figure}[!bht]
\centering
  \includegraphics[width=\textwidth]{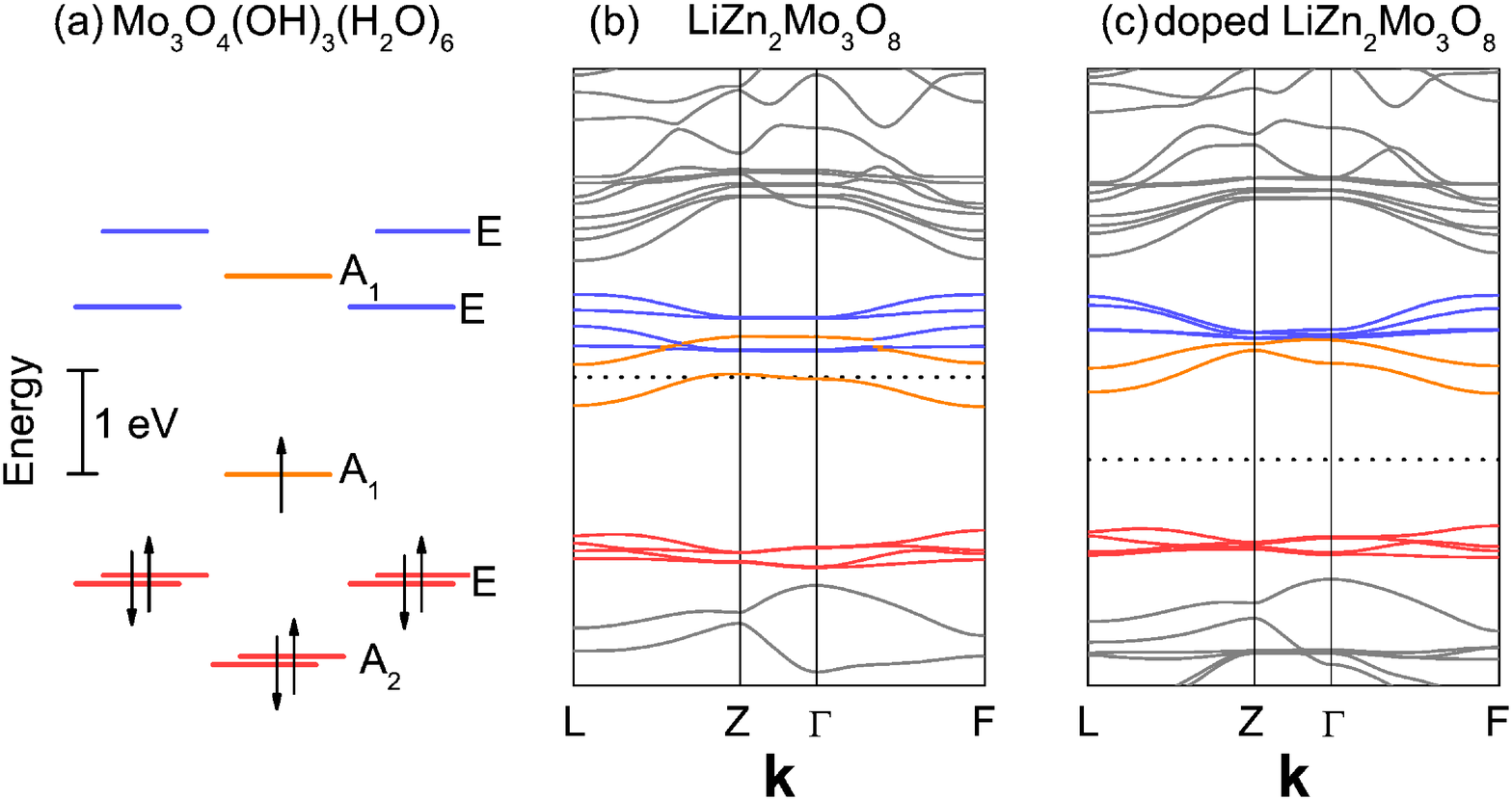}
   \caption{Electronic structure calculations using density functional theory. Plots are offest to lineup the centroid of frontier orbitals. Band structures are calculated in the $R\bar{3}m$ rhombohedral setting, with two \MoOC clusters per unit cell. Dashed lines are respective fermi energies. (a) A molecular orbital diagram calculated for \MoOH, an electronic analogue to a \MoOC cluster in \LiZn, taken from ref.~\cite{sheckelton_possible_2012}. This shows a \SOneHalf arising in each molecule. This is consistent with (b), A band structure calculation on \LiZn. Here, a Hubbard U of 2 eV was chosen for the calculations in (b) and (c), doped \LiZn. As can be seen in (b), the band corresponding to the highest occupied molecular orbital in (a), once doped via the removal of Li in (c) shifts up in energy relative to the valence and conduction bands. While it is unphysical to apply a Hubbard U to a closed shell system in (c), the calculations were performed this way so that the undoped and doped band structures can be directly compared. These calculations suggest that upon doping, the energy level of the resulting hole increases in energy, thus favoring a form of Anderson Localization and an increase in the bandgap.}
  \label{fig:Band}
\end{figure}

\FloatBarrier

\newpage
\section{\LiZn Crystallographic and Local Structure}
\begin{figure*}[!bth]
\centering
  \includegraphics[width=\textwidth]{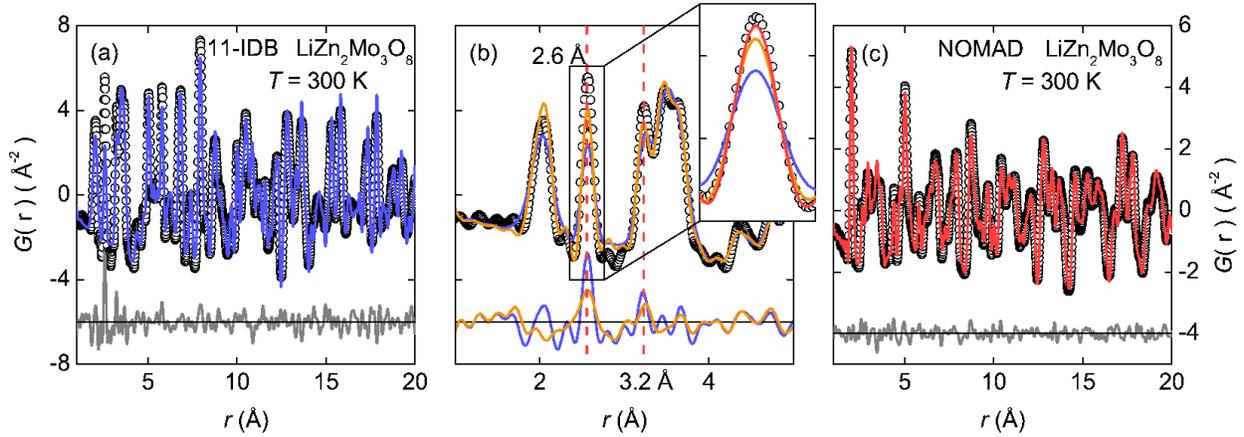}
  \caption{(a) Least squares fit to X-ray PDF data on \LiZn using the average $R\bar{3}m$ structure at \textit{T}~=~300\,K. The fit is consistent with the calculated structure from the combined SXRD/NPD rietveld refinement, except for the large deviation from the data at low \textit{r}, which is due to correlated motion effects. (b) The same refinement at \textit{r}~$\leq 5$\,\AA{} (blue line and difference curve) and a least squares fit to an overall 1/$r$ correlated motion factor (orange lines). The two dashed red lines correspond to intra- (2.6\,\AA{}) and inter- (3.2\,\AA{}) Mo-Mo bond distances. Despite the refinement of the correlated motion factor in (b), the low \textit{r} refinement is still poorly fit to the data. The inset shows the two fits (blue and orange lines) for the peak corresponding to the Mo-Mo intra-cluster peak at 2.6\,$\textrm{\AA{}}$. In addition, the red line is the original fit to the whole PDF [as in (a), the blue line] but with all parameters fixed and only the linear correlated motion parameter refined for the intra-cluster Mo-Mo peak. This indicates the correlated motion factor varies for different parts of the structure, namely \MoOL and the Li/Zn interlayers. This is expected, chemically, from the strong Mo-Mo bonding that is present. Note that such low $r$ peak sharpening is inconsistent with a deviation of the local structure from trigonal symmetry, which would tend to broaden features in the PDF. (c) Least squares fit to neutron PDF data at \textit{T}~=~300\,K. As opposed to the X-ray fit in (a), a single overall correlated motion factor results in a good fit to the data, suggesting that the correlated motion of the \MoOL is on a slower timescale than the interaction timescale of the neutrons.}
  \label{fig:Xray_PDFs}
\end{figure*}

\begin{figure}[!bth]
\centering
  \includegraphics[width=4in]{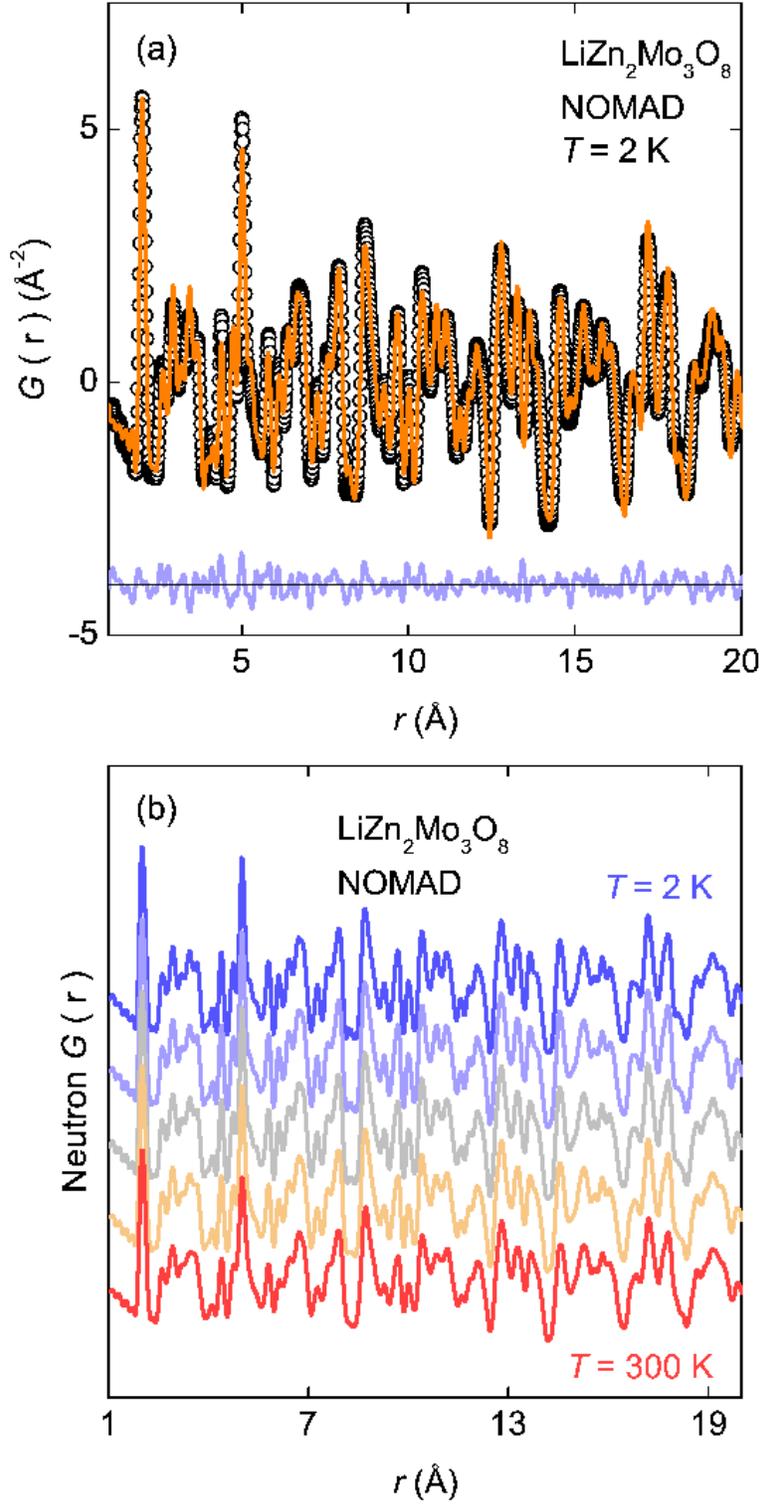}
  \caption{(a) Neutron PDF and corresponding least squares fit of undoped \LiZn at \textit{T}~=~2\,K, using refined average structure in $R\bar{3}m$. The average structure yields a good fit to the PDF over all ranges of $r$ using an overall $1/r$ correlated motion factor as in Figure~\ref{fig:Xray_PDFs}(c). (b) Neutron PDFs of \LiZn at various temperatures are surprisingly similar, with differences attributed only to the expected changes in thermal broadening, indicative of the lack of formation of any local structural distortions upon cooling down to \textit{T}~=~2\,K.}
  \label{fig:Neutron_PDFs}
\end{figure}

\FloatBarrier

\begin{figure}[!hbt]
\centering
  \includegraphics[width=4in]{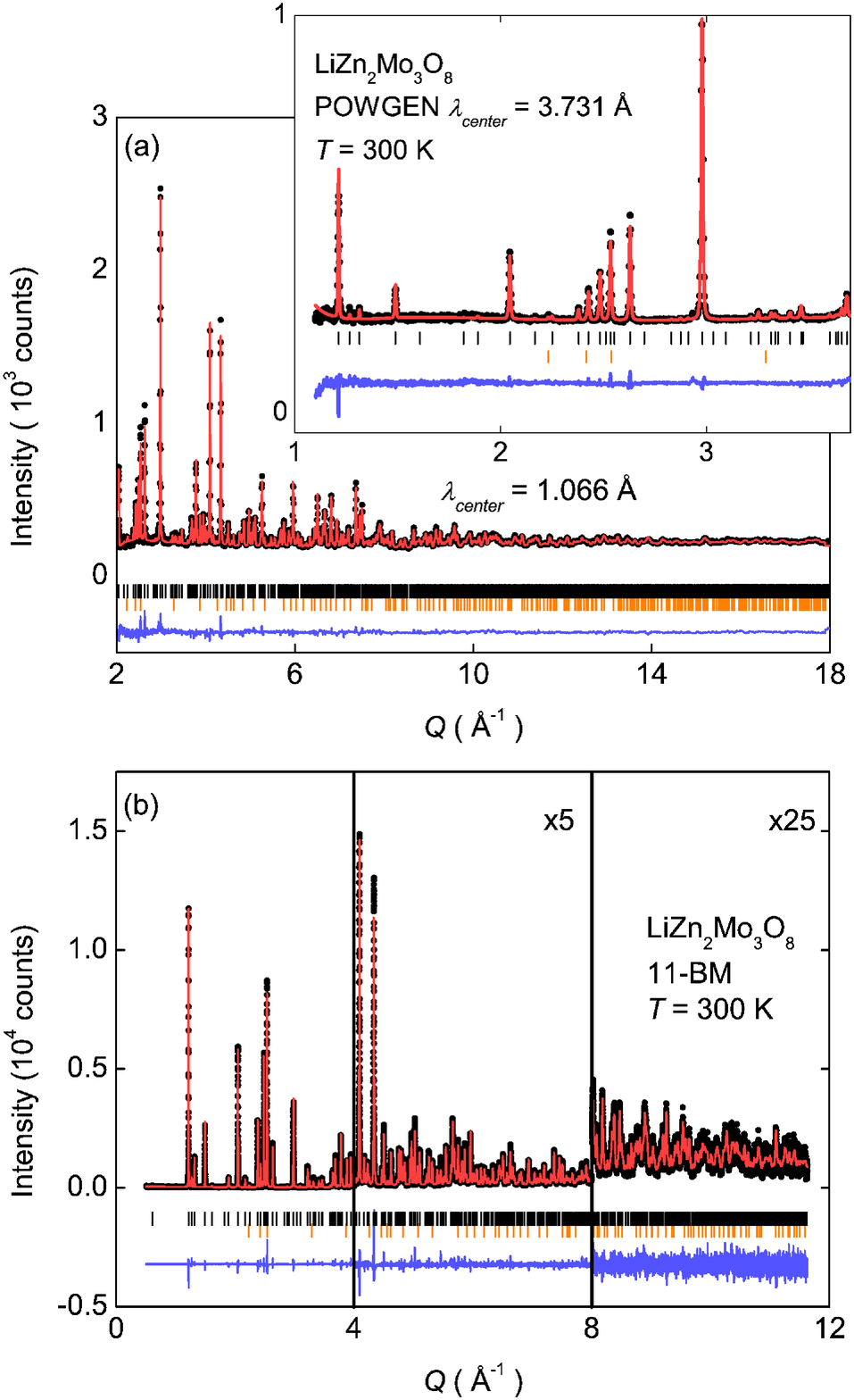}
  \caption{A joint Rietveld refinement to NPD (a) and SXRD (b) datasets on undoped \LiZn. Black dots are observed data, red line is Rietveld fit, blue line is the difference, black tick lines correspond to \LiZn reflections in the $R\bar{3}m$ spacegroup, and orange lines are ZnO impurity reflections. Results are summarized in table~\ref{tab:NX_Riet} and table~\ref{tab:Riet2}.}
  \label{fig:XN_fit}
\end{figure}

\begin{table}[!htb]
 \small
  \caption{Combined SXRD/Neutron refinement of \LiZn at \textit{T} = 300 K}
  \label{tab:NX_Riet}
   \begin{tabular*}{\columnwidth}{@{\extracolsep{\fill}}lr}
    \hline 
    Chemical formula & Li${_{1.0(1)}}$Zn$_{1.8(1)}$Mo$_3$O$_8$ \\ 
    \hline
    Space group & $R\bar{3}m$\\
    a(\AA{}) & 5.80163(3)\\
    c(\AA{}) & 31.0738(2)\\
    Z & 6\\
    Global weighted $\chi^2$ & 6.870 \\
    \hline 
   \end{tabular*}
\end{table}

\begin{table*}[tbh!]
 \small
  \caption{Atomic parameters of Li$_{1.0(1)}$Zn$_{1.8(1)}$Mo$_3$O$_8$\xspace from joint SXRD/Neutron Rietveld refinement at \textit{T}~=~300\,K}
  \label{tab:Riet2}
   \begin{tabular*}{\textwidth}{@{\extracolsep{\fill}}clllcll}
    \hline 
    \bf{atom} & x & y & z & Wyckoff position & Occupancy & U$_{iso}$ \\ 
    \hline
     \textbf{Mo-1} & 0.18504(8) & 0.81496(8) & 0.08393(4) & 18\textit{h} & 1.0000 & 0.0016(2) \\      
     \textbf{O-1} & 0.8445(2) & 0.1555(2) & 0.04850(6) & 18\textit{h} & 1.0000 & 0.0017(3)\\      
     \textbf{O-2} & 0.4920(2) & 0.5080(2) & 0.12438(7) & 18\textit{h} & 1.0000 & 0.0047(4)\\      
     \textbf{O-3} & 0 & 0 & 0.1185(1) & 6\textit{c} & 1.0000 & 0.0054(7)\\      
     \textbf{O-4} & 0 & 0 & 0.3715(1) & 6\textit{c} & 1.0000 & 0.0053(6)\\     
     \textbf{Zn-1} & 1/3 & 2/3 & -0.64176(7) & 6\textit{c} & 0.879(6) & 0.0038(4)\\    
     \textbf{Li-1} & 1/3 & 2/3 & -0.64176(7) & 6\textit{c} & 0.00(4) & 0.0038(4)\\
     \textbf{Zn-2} & 0 & 0 & 0.1813(1) & 6\textit{c} & 0.679(5) & 0.0038(4)\\    
     \textbf{Li-2} & 0 & 0 & 0.1813(1) & 6\textit{c} & 0.22(4) & 0.0048(4)\\
     \textbf{Zn-3} & 0 & 0 & 0 & 3\textit{a} & 0.265(7) & 0.0038(4)\\    
     \textbf{Li-3} & 0 & 0 & 0 & 3\textit{a} & 0.58(6) & 0.0038(4)\\
     \textbf{Zn-4} & 0 & 0 & 0.5070(8) & 6\textit{c} & 0.065(4) & 0.0038(4)\\ 
     \textbf{Li-4} & 0 & 0 & 0.5070(8) & 6\textit{c} & 0.43(3) & 0.0038(4)\\
     \textbf{Li-5} & 2/3 & 1/3 & 0.08392(4) & 6\textit{c} & 0.09(3) & 0.0038(4)\\   
    \hline 
   \end{tabular*}
\end{table*}

\FloatBarrier
\section{Synchrotron X-ray analysis of \LiZnx ($2x < 0.4$)} 

\begin{table*}[htb!]
\small
 \caption{\ Unit cell and fit statistics from \LiZnx Rietveld refinements of 11-BM data at \textit{T}~=~6\,K.}
  \label{tab:Riet}
   \begin{tabular*}{\textwidth}{@{\extracolsep{\fill}}cllllclc}
    \hline 
    Sample (2x) & $a$(\AA{}) & $c$(\AA{}) & $c/a$ & $R_{wp}$ & LeBail $R_{wp}$ & ${\chi}^2$ & LeBail $\chi^2$ \\ 
    \hline
    0.05 & 5.78148(2) & 31.007(1) & 5.36329(3) & 11.96 & 10.09 & 5.48 & 3.96\\
    0.10 & 5.78097(2) & 30.997(1) & 5.36198(3) & 12.12 & 10.33 & 4.33 & 3.20\\
    0.15 & 5.780(5) & 30.98(3) & 5.3605(7) & 11.69 & 10.17 & 4.16 & 3.20\\
    0.20 & 5.7794(2) & 30.972(1) & 5.35910(3) & 11.68 & 10.14 & 3.92 & 3.03\\
    0.25 & 5.7777(8) & 30.954(4) & 5.3575(1) & 11.94 & 10.38 & 3.20 & 2.47\\
    0.30 & 5.777(2) & 30.947(1) & 5.3562(2) & 11.90 & 10.44 & 5.62 & 4.40\\
    0.40 & 5.7763(1) & 30.921(1) & 5.35318(3) & 11.90 & 10.35 & 4.75 & 3.65\\
    \hline 
   \end{tabular*}
\end{table*}

\begin{figure}[!hbt]
\centering
  \includegraphics[width=4in]{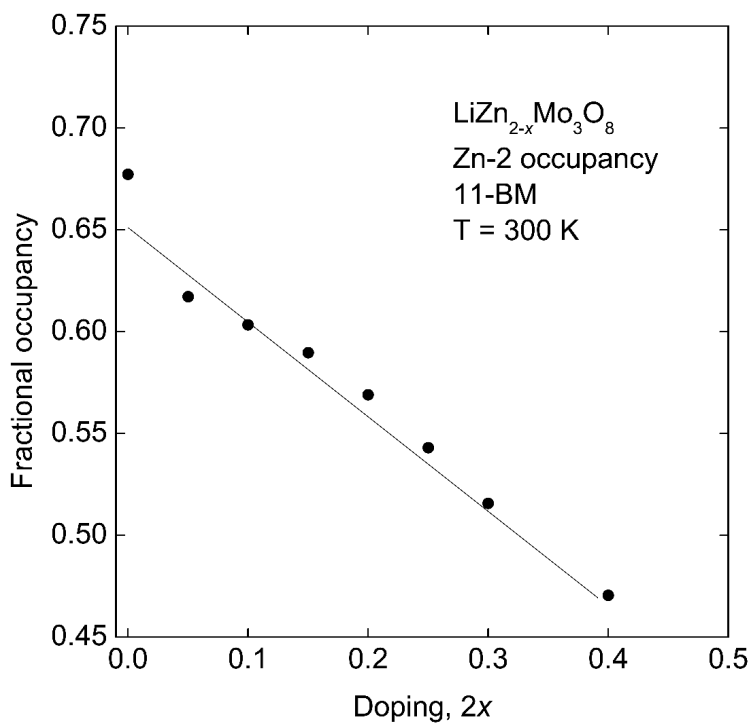}
  \caption{Occupation of the Zn-2 site obtained from Rietveld refinements of SXRD data. The line is a linear fit to the $x > 0$ data. The slope of this line, -0.45, shows that the bulk of the deintercalated zinc (90\%) is removed from from the Zn-2 site. }
  \label{fig:Zn2_occ}
\end{figure}

\FloatBarrier

\bibliography{./references/Doping}

\end{document}